\documentclass[preprint,12pt]{elsarticle}

\usepackage{amssymb}
\newcommand\rf[1]{(\ref{eq:#1})}
\newcommand\lab[1]{\label{eq:#1}}
\newcommand\nonu{\nonumber}
\newcommand\br{\begin{eqnarray}}
\newcommand\er{\end{eqnarray}}
\newcommand\be{\begin{equation}}
\newcommand\ee{\end{equation}}

\newcommand\lb{\lbrack}
\newcommand\rb{\rbrack}

\newcommand\llb{\left\lbrack}
\newcommand\rrb{\right\rbrack}

\renewcommand\({\left(}
\renewcommand\){\right)}
\newcommand\bc{\begin{center}}
\newcommand\ec{\end{center}}

\renewcommand\a{\alpha}
\renewcommand\b{\beta}

\newcommand\vareps{\varepsilon}

\newcommand\G{\Gamma}

\newcommand\h{\frac{1}{2}}
\renewcommand\k{\kappa}
\renewcommand\l{\lambda}
\renewcommand\L{\Lambda}
\newcommand\m{\mu}
\newcommand\n{\nu}

\newcommand\vp{\varphi}
\renewcommand\P{\Phi}
\newcommand\pa{\partial}

\renewcommand\th{\theta}

\newcommand\cE{{\mathcal E}}

\newcommand\cJ{{\mathcal J}}

\newcommand\cV{{\mathcal V}}

\newcommand{\ct}[1]{\cite{#1}}
\newcommand{\bib}[1]{\bibitem{#1}}

\newcommand\PRL[3]{\textsl{Phys. Rev. Lett.} \textbf{#1} (#2) #3}
\newcommand\NPB[3]{\textsl{Nucl. Phys.} \textbf{B#1} (#2) #3}

\newcommand\PRD[3]{\textsl{Phys. Rev.} \textbf{D#1} (#2) #3}

\newcommand\PLB[3]{\textsl{Phys. Lett.} \textbf{#1B} (#2) #3}
\newcommand\CQG[3]{\textsl{Class. Quantum Grav.} \textbf{#1} (#2) #3}

\newcommand\AoP[3]{\textsl{Ann. of Phys.} \textbf{#1} (#2) #3}

\newcommand\IJMPA[3]{\textsl{Int. J. Mod. Phys.} \textbf{A#1} (#2) #3}

\newcommand\MPLA[3]{\textsl{Mod. Phys. Lett.} \textbf{A#1} (#2) #3}

\newcommand\xdot{\stackrel{.}{x}}

\begin{document}

\begin{frontmatter}

%% Title, authors and addresses

%% use the tnoteref command within \title for footnotes;
%% use the tnotetext command for theassociated footnote;
%% use the fnref command within \author or \address for footnotes;
%% use the fntext command for theassociated footnote;
%% use the corref command within \author for corresponding author footnotes;
%% use the cortext command for theassociated footnote;
%% use the ead command for the email address,
%% and the form \ead[url] for the home page:
%% \title{Title\tnoteref{label1}}
%% \tnotetext[label1]{}
%% \author{Name\corref{cor1}\fnref{label2}}
%% \ead{email address}
%% \ead[url]{home page}
%% \fntext[label2]{}
%% \cortext[cor1]{}
%% \address{Address\fnref{label3}}
%% \fntext[label3]{}

\title{Asymptotically de Sitter and anti-de Sitter Black Holes with
Confining Electric Potential}

%% use optional labels to link authors explicitly to addresses:
%% \author[label1,label2]{}
%% \address[label1]{}
%% \address[label2]{}

\author[BGU]{Eduardo Guendelman\corref{cor1}}
\ead{guendel@bgu.ac.il}
\ead[url]{http://eduardo.hostoi.com}
\cortext[cor1]{Corresponding author -- tel. +972-8-647-2508, fax +972-8-647-2904.}
% \author[BGU]{Alexander Kaganovich\corref{cor1}}
\author[BGU]{Alexander Kaganovich}
\ead{alexk@bgu.ac.il}
\ead[url]{http://profiler.bgu.ac.il/frontoffice/ShowUser.aspx?id=1249}
\address[BGU]{Department of Physics, Ben-Gurion University of the Negev,
P.O.Box 653, IL-84105 ~Beer-Sheva, Israel}

\author[INRNE]{Emil Nissimov}
\ead{nissimov@inrne.bas.bg}
\ead[url]{http://theo.inrne.bas.bg/~nissimov/}
\author[INRNE]{Svetlana Pacheva}
\ead{svetlana@inrne.bas.bg}
\ead[url]{http://theo.inrne.bas.bg/~svetlana/}
\address[INRNE]{Institute for Nuclear Research and Nuclear Energy, Bulgarian Academy
of Sciences, Boul. Tsarigradsko Chausee 72, BG-1784 ~Sofia, Bulgaria}

\begin{abstract}
We study gravity interacting with a special kind of QCD-inspired nonlinear
gauge field system which earlier was shown to yield confinement-type effective 
potential (the ``Cornell potential'') between charged fermions (``quarks'')
in flat space-time. 
We find new static spherically symmetric solutions generalizing the usual 
Reissner-Nordstr{\"o}m-de-Sitter and Reissner-Nordstr{\"o}m-{\em anti}-de-Sitter
black holes with the following additional properties: 
(i) appearance of a constant radial electric field (in addition to the Coulomb one);
(ii) novel mechanism of {\em dynamical generation} of cosmological constant
through the non-Maxwell gauge field dynamics; 
(iii) appearance of confining-type effective potential in charged test particle 
dynamics in the above black hole backgrounds.
\end{abstract}

\begin{keyword}
%% keywords here, in the form: keyword \sep keyword
black holes in (anti) de Sitter spaces \sep dynamically generated cosmological 
constant \sep QCD-inspired confining potential

%% PACS codes here, in the form: \PACS code \sep code
\PACS 11.25.-w \sep 04.70.-s \sep 04.50.+h
%% MSC codes here, in the form: \MSC code \sep code
%% or \MSC[2008] code \sep code (2000 is the default)
\end{keyword}

\end{frontmatter}

%% \linenumbers

%% main text
%%%%%%%%%%%%%%%%%%%%%%%%%%%%%%%%%%%%%%%%%%%%%%%%%%%%%%%%%%%%%%%%%%%%%%%%%%%%%%%%%%
\section{Introduction}
\label{intro}

% Attempts to exploit non-Maxwell nonlinear effective gauge field models to
% describe confinement phenomena have long history \ct{tHooft-03}.

% The main motivation of the present work stems from the previous studies \ct{GG}
% of special type of non-Maxwell nonlinear gauge field theory in flat Minkowski 
% space-time:

It has been shown by `t Hooft \ct{tHooft-03} that in any effective quantum theory,
which is able to describe linear confinement phenomena, the energy density of
electrostatic field configurations should be a linear function of the
electric displacement field in the infrared region %% ADD
due to appropriate infrared counterterms.%% ADD
The simplest way to achieve this in Minkowski space-time is by considering 
a square root of the field strength squared, in addition to the standard Maxwell term,
leading to a very peculiar non-Maxwell nonlinear effective gauge field model \ct{GG}:
\br
S = \int d^4 x L(F^2) \quad ,\quad
L(F^2) = -\frac{1}{4} F^2 - \frac{f}{2} \sqrt{-F^2} \; ,
\lab{GG} \\
F^2 \equiv F_{\m\n} F^{\m\n} \quad ,\quad 
F_{\m\n} = \pa_\m A_\n - \pa_\n A_\m  \; ,
\nonu
\er
with $f$ being a positive coupling constant. %% ADD
It has been shown in first three refs.\ct{GG} that the square root of the Maxwell term naturally 
arises as a result of spontaneous breakdown of scale symmetry of the original
scale-invariant Maxwell theory with $f$ appearing as an integration constant 
responsible for the latter spontaneous breakdown. %% ADD
The model \rf{GG} produces a confining 
effective potential $V(r) = - \frac{\a}{r} + \b r$ (Coulomb plus 
linear one) which is of the form of the well-known ``Cornell'' potential 
\ct{cornell-potential} in quantum chromodynamics (QCD).
For static field configurations the model \rf{GG} yields
the following electric displacement field
$\vec{D} = \vec{E} - \frac{f}{\sqrt{2}}\frac{\vec{E}}{|\vec{E}|}$. The pertinent energy
density turns out to be 
(there is {\em no} contribution from the square-root term in \rf{GG})
$\h \vec{E}^2 = \h |\vec{D}|^2 + \frac{f}{\sqrt{2}} |\vec{D}| + \frac{1}{4} f^2$, 
so that it indeed contains a term linear w.r.t. $|\vec{D}|$.

It is crucial to stress that the Lagrangian $L(F^2)$ \rf{GG} contains both the 
usual Maxwell term as well as a non-analytic function of $F^2$ and thus it
is a {\em non-standard} form of nonlinear electrodynamics. In this way it is 
significantly different from the original ``square root'' Lagrangian 
$- \frac{f}{2}\sqrt{F^2}$ first proposed by Nielsen and Olesen \ct{N-O} to
describe string dynamics. Moreover, it is important that the square root term in \rf{GG}
is in the ``electrically'' dominated form ($\sqrt{-F^2}$) as opposed to the 
``magnetically'' dominated Nielsen-Olesen form ($\sqrt{F^2}$).

Let us remark that one could start with the non-Abelian version of the action \rf{GG}.
Since we will be interested in static spherically symmetric solutions, the
non-Abelian theory effectively reduces to an Abelian one as pointed out in
the first ref.\ct{GG}.

Our main goal in the present note is to study possible new effects by
coupling the confining potential generating nonlinear gauge field system \rf{GG} 
to gravity. We find:

(i) appearance of a constant radial electric field (in addition to the Coulomb one)
in charged black holes within Reissner-Nordstr{\"o}m-de-Sitter % (RN-dS) 
and/or Reissner-Nordstr{\"o}m-{\em anti}-de-Sitter % (RN-AdS) 
space-times as well as in electrically neutral black holes with Schwarzschild-de-Sitter 
% (S-dS) 
and/or Schwarzschild-{\em anti}-de-Sitter % (S-AdS) 
geometry;

(ii) novel mechanism of {\em dynamical generation} of cosmological constant
through the non-Maxwell gauge field dynamics of the nonlinear action $L(F^2)$
\rf{GG}; 

(iii) appearance of confining-type effective potential in charged test particle 
dynamics in the above black hole backgrounds.

%%%%%%%%%%%%%%%%%%%%%%%%%%%%%%%%%%%%%%%%%%%%%%%%%%%%%%%%%%%%%%%%%%%%%%%%%%%%%%%%%%
\section{Lagrangian Formulation. Spherically Symmetric Solutions}
\label{lagrange}

We will consider the simplest coupling of the nonlinear gauge field system
\rf{GG} to gravity described by the action (we use units with Newton
constant $G_N=1$):
\be
S = \int d^4 x \sqrt{-g} \Bigl\lb \frac{R(g)}{16\pi} 
- \frac{1}{4} F^2 - \frac{f}{2} \sqrt{-F^2}\Bigr\rb  \quad ,\quad
F^2 \equiv F_{\k\l} F_{\m\n} g^{\k\m} g^{\l\n}  \; ,
% \quad ,\quad F_{\m\n} = \pa_\m A_\n - \pa_\n A_\m
\lab{gravity+GG}
\ee
where $R(g)$ is the scalar curvature of the space-time metric
$g_{\m\n}$ and $g \equiv \det\Vert g_{\m\n}\Vert$.
It is important to stress that for the time being we will {\em not}
introduce any bare cosmological constant term.

The energy-momentum tensor $T^{(F)}_{\m\n}$ of the nonlinear gauge field, which
appears in the pertinent equations of motion:
\br
R_{\m\n} - \h g_{\m\n} R = 8\pi T^{(F)}_{\m\n} \; ,
\lab{einstein-eqs}\\
\pa_\n \(\sqrt{-g}\Bigl( 1 -\frac{f}{\sqrt{-F^2}}\Bigr) F_{\k\l} g^{\m\k} g^{\n\l}\)=0
\; ,
\lab{GG-eqs}
\er
% where $L^{\pr} (F^2)$ denotes derivative w.r.t. $F^2$ of the function $L(F^2)$ in 
% \rf{GG}, 
is explicitly given by:
\be
% T^{(F)}_{\m\n} = L(F^2) g_{\m\n} - 4 L^{\pr}(F^2) F_{\m\k} F_{\n\l} g^{\k\l}
T^{(F)}_{\m\n} = \Bigl( 1 - \frac{f}{\sqrt{-F^2}}\Bigr) F_{\m\k} F_{\n\l} g^{\k\l}
- \frac{1}{4} \Bigl( F^2 + 2f\sqrt{-F^2}\Bigr) g_{\m\n} \; .
\lab{stress-tensor-F}
\ee
%
%%%%%%%%%%%%%%%%%%%%%%%%%%%%%%%%%%%%%%%%%%%%%%%%%%%%%%%%%%%%%%%%%%%%%%%%%%%%%%%%%%
% \section{Static Spherically Symmetric Solutions}
% \label{solution}
We will look for static spherically symmetric solutions of the system
\rf{einstein-eqs}--\rf{stress-tensor-F}:
\br
ds^2 = - A(r) dt^2 + \frac{dr^2}{A(r)} + r^2 \bigl(d\th^2 + \sin^2 \th d\vp^2\bigr)
\; ,
\lab{spherical-static} \\
F_{\m\n} = 0 \;\; \mathrm{for}\; (\m,\n)\neq (0,r) \quad ,\quad
F_{0r} = F_{0r} (r) \; .
\lab{electr-static}
\er
In this case the gauge field equations of motion \rf{GG-eqs} become:
\be
\pa_r \( r^2 \Bigl(F_{0r} - \frac{\vareps_F f}{\sqrt{2}}\Bigr)\) = 0
\quad ,\quad \vareps_F \equiv \mathrm{sign}(F_{0r}) \; ,
\lab{GG-eqs-0}
\ee
whose solution reads:
\be
F_{0r} = \frac{\vareps_F f}{\sqrt{2}} + \frac{Q}{\sqrt{4\pi}\, r^2} 
\quad ,\quad \vareps_F = \mathrm{sign}(Q) \; .
\lab{cornell-sol}
\ee
Again, as in the flat space-time case \rf{GG}, the electric field contain a radial
constant piece $\vareps_F f/\sqrt{2}$ alongside with the Coulomb term.

Further, it has been shown in ref.\ct{Ed-Rab} that for static spherically symmetric
metrics \rf{spherical-static} with the associated energy-momentum tensor
obeying the condition $T^0_0 = T^r_r$, which is fulfilled in the present case
\rf{electr-static}, it is sufficient to solve Einstein equations:
\br
R^0_0 = 8\pi (T^0_0 - \h T^\l_\l) \quad \mathrm{where} \;\;
R^0_0 = - \frac{1}{2r^2} \pa_r \bigl( r^2 \pa_r A\bigr) \; ,
% \lab{Ed-Rab-1} \\
\nonu \\
R^\th_\th = - 8\pi T^0_0 \quad \mathrm{where} \;\;
R^\th_\th = - \frac{1}{r^2} (A-1) - \frac{1}{r}\pa_r A \; .
\lab{Ed-Rab}
\er
% (in $D=4$ space-time dimensions):\
In the case under consideration the r.h.s. of the first Einstein equation \rf{Ed-Rab} 
with the energy-momentum tensor \rf{stress-tensor-F} becomes:
\be
8\pi \Bigl({T^{(F)}}^0_0 - \h {T^{(F)}}^\l_\l)\Bigr) =
% - 4\pi \(\frac{f}{\sqrt{2}} + \frac{Q}{\sqrt{4\pi}\, r^2}\)^2
% + 4\pi\sqrt{2}f\,\(\frac{f}{\sqrt{2}} + \frac{Q}{\sqrt{4\pi}\, r^2}\)
-4\pi \(\frac{Q^2}{4\pi\, r^4} - \h f^2\)
\lab{einstein-eqs-rhs}
\ee
taking into account \rf{electr-static} and \rf{cornell-sol}. Interestingly
enough, there are no cross terms in \rf{einstein-eqs-rhs} between the Coulomb and 
constant electric parts.

The solution of \rf{Ed-Rab} with \rf{einstein-eqs-rhs} yields:
\be
A(r) = 1 - \sqrt{8\pi}|Q|f - \frac{2m}{r} + \frac{Q^2}{r^2} - \frac{2\pi f^2}{3} r^2 \; .
\lab{RN-dS+const-electr}
\ee
In other words the solution given by \rf{spherical-static}, \rf{RN-dS+const-electr}
and \rf{cornell-sol} describes a black hole with:
\begin{itemize}
\item
Reissner-Nordstr{\"o}m-de-Sitter space-time geometry \rf{RN-dS+const-electr};
\item
additional global constant radial electric field in \rf{cornell-sol} apart from the 
usual Coulomb one;
\item
{\em dynamically generated} effective cosmological constant in \rf{RN-dS+const-electr}
(let us recall that there was {\em no} bare cosmological constant in \rf{gravity+GG}):
\be
\L_{\mathrm{eff}} = 2\pi f^2 \; .
\lab{CC-eff}
\ee
\end{itemize}

In particular, when $Q=0$ we obtain electrically neutral black hole with
Schwarzschild-de-Sitter geometry:
\be
A(r) = 1 - \frac{2m}{r} - \frac{2\pi f^2}{3} r^2 \: ,
\lab{S-dS+const-electr}
\ee
where the cosmological constant \rf{CC-eff} is {\em dynamically generated},
and with additional global constant radial electric field:
\be
F_{0r} = \vareps_F f/\sqrt{2} \; .
\lab{const-electr}
\ee

%%%%%%%%%%%%%%%%%%%%%%%%%%%%%%%%%%%%%%%%%%%%%%%%%%%%%%%%%%%%%%%%%%%%%%%%%%%%%%%%%%
\section{Bare Negative Cosmological Constant versus Induced Cosmological Constant}
\label{versus}
Let us now introduce in \rf{gravity+GG} from the very beginning a negative bare 
cosmological constant $\L = - |\L|$:
\be
S = \int d^4 x \sqrt{-g} \Bigl\lb \frac{1}{16\pi} \bigl( R(g)- 2\L\bigr)
- \frac{1}{4} F^2 - \frac{f}{2} \sqrt{-F^2}\Bigr\rb \; .
\lab{gravity+GG+CC}
\ee
Then the corresponding static spherically symmetric solution is given by
\rf{cornell-sol} and \rf{spherical-static} with:
% \br
% F_{0r} = \frac{f}{\sqrt{2}} + \frac{Q}{\sqrt{4\pi}r^2} 
% \lab{const-electr-1} \\
% ds^2 = - A(r) dt^2 + \frac{dr^2}{A(r)} + r^2 \bigl(d\th^2 + \sin^2 \th d\vp^2\bigr)
% \nonu \\
% \er
\be
A(r) = 1 - \sqrt{8\pi}|Q|f - \frac{2m}{r} + \frac{Q^2}{r^2} + 
\frac{1}{3} \bigl(|\L| - 2\pi f^2) r^2 \; .
\lab{RN-AdS+const-electr} 
\ee
Thus, we find also black hole solution with Reissner-Nordstr{\"o}m-{\em anti-de-Sitter} 
geometry \rf{RN-AdS+const-electr} and with additional global constant electric field
\rf{cornell-sol} provided the full effective cosmological constant (bare one 
plus {\em dynamically induced} one) satisfies:
\be
\L_{\mathrm{eff}} = - |\L| + 2\pi f^2 < 0 \quad ,\quad \textsl{i.e.} \;\;
|\L| > 2\pi f^2 \;.
\lab{CC-eff-1}
\ee
On the other hand, if $|\L| < 2\pi f^2$, \textsl{i.e.} 
$\L_{\mathrm{eff}} = 2\pi f^2 - |\L| > 0$, the solution
\rf{RN-AdS+const-electr} describes asymptotically de Sitter black hole
{\em in spite} of the presence of negative bare cosmological constant $\L$. In the
special case $|\L| = 2\pi f^2$ the dynamically induced cosmological constant
completely cancels the effect of the negative bare cosmological constant
and the resulting solution describes an asymptotically flat Reissner-Nordstr{\"o}m-like 
black hole ($A(r) = 1  - \sqrt{8\pi}|Q|f - \frac{2m}{r} + \frac{Q^2}{r^2}$)
with additional global constant radial electric field \rf{const-electr}.

In particular, when $Q=0$ the solution \rf{RN-AdS+const-electr} reduces to 
electrically neutral black hole with Schwarzschild-{\em anti}-de-Sitter geometry 
for $|\L| > 2\pi f^2$:
\be
A(r) = 1 - \frac{2m}{r} + \frac{1}{3} \bigl(|\L|- 2\pi f^2) r^2 \; ,
\lab{S-AdS+const-electr}
\ee
or electrically neutral black hole with 
Schwarzschild-de-Sitter geometry for $|\L| < 2\pi f^2$.
In both cases above an additional global constant radial electric field
\rf{const-electr} is present.

In the special case $|\L| = 2\pi f^2$ and $Q=0$ we obtain asymptotically flat 
Schwarzschild black hole ($A(r) = 1 - \frac{2m}{r}$) with additional global 
constant radial electric field \rf{const-electr}, \textsl{i.e.},
$\L_{\mathrm{eff}}=0$ in spite of the presence of 
the negative bare cosmological constant in the gravity-gauge-field action 
\rf{gravity+GG+CC}.
%%%%%%%%%%%%%%%%%%%%%%%%%%%%%%%%%%%%%%%%%%%%%%%%%%%%%%%%%%%%%%%%%%%%%%%%%
\section{Charged Test Particle Dynamics}
\label{particle}

Let us now briefly discuss the dynamics of a test particle with mass $m_0$
and electric charge $q_0$ in the above black hole backgrounds -- 
\rf{spherical-static} with \rf{RN-dS+const-electr} 
and \rf{cornell-sol} or \rf{spherical-static} with \rf{RN-AdS+const-electr} 
and \rf{cornell-sol}. It is given by the standard reparametrization
invariant point-particle action:
\be
S_{\mathrm{particle}} = \int d\l \llb \frac{1}{2e}g_{\m\n}(x) \xdot^\m \xdot^\n
-\h e m_0^2 - q_0 \xdot^\m\!\! A_\m (x)\rrb \; ,
\lab{test-particle}
\ee
where $e$ denotes the world-line ``einbein''. The standard treatment, using 
energy $\cE$ and angular momentum $\cJ$ conservation in the static spherically 
symmetric background under consideration and replacing the arbitrary world-line 
parameter $\l$ with the particle proper-time parameter $s$ via 
$\frac{ds}{d\l} = e m_0$, yields the radial motion equation:
\be
\(\frac{dr}{ds}\)^2 + \cV_{\mathrm{eff}}(r) = \frac{\cE^2}{m_0^2} \; ,
\lab{radial-eq}
\ee
\br
\cV_{\mathrm{eff}}(r) \equiv A(r) \Bigl( 1 + \frac{\cJ^2}{m_0 r^2}\Bigr)
- \frac{q_0^2}{m_0^2}\, r^2 
\Bigl(\frac{\vareps_F f}{\sqrt{2}} - \frac{Q}{\sqrt{4\pi}\, r^2}\Bigl)^2
\nonu \\
- 2\, \frac{\cE q_0}{m_0^2}\, r 
\Bigl(\frac{\vareps_F f}{\sqrt{2}} - \frac{Q}{\sqrt{4\pi}\, r^2}\Bigl) \; ,
\lab{V-eff}
\er
with $A(r)$ as in \rf{RN-dS+const-electr} or \rf{RN-AdS+const-electr}.

Taking for simplicity $Q=0$ (neutral black hole background) and $\cJ=0$
(zero impact parameter -- purely radial motion) the ``effective'' potential
\rf{V-eff} becomes:
\be
\cV^{(0)}_{\mathrm{eff}}(r) = 1 - \frac{2m}{r} + 
\Bigl(\frac{1}{3}\bigl(|\L| - 2\pi f^2\bigr) - \frac{q_0^2 f^2}{2m_0^2}\Bigr)\, r^2
- \frac{\sqrt{2}\cE q_0 \vareps_F f}{m_0^2}\, r \; .
\lab{V-eff-0}
\ee
In a Schwarzschild-{\em anti}-de-Sitter black hole \rf{S-AdS+const-electr}
with constant radial electric field \rf{const-electr}, 
for the special value of the ratio of the test particle parameters
$q_0^2/m_0^2 = 2/3 \bigl(|\L|/f^2 - 2\pi\bigr)$ the term quadratic w.r.t.
$r$ in \rf{V-eff-0} vanishes and % $\cV^{(0)}_{\mathrm{eff}}$ 
the latter acquires the form of a QCD-like (``Cornell''-type \ct{cornell-potential})
confining-type potential (provided 
$q_0 \vareps_F < 0$ with $\vareps_F$ as in \rf{GG-eqs-0}):
\be
\cV^{(0)}_{\mathrm{eff}}(r) = 
% 1 - \frac{2m}{r} - \frac{\sqrt{2}\cE q_0 \vareps_F f}{m_0^2}\, r \; .
1 - \frac{2m}{r} + \frac{\sqrt{2}\cE |q_0| f}{m_0^2}\, r \; .
\lab{V-eff-cornell}
\ee
%%%%%%%%%%%%%%%%%%%%%%%%%%%%   %%%%%%%%%%%%%%%%%%%%%%%%%%%%

Let us particularly stress that the ``Cornell''-type confining potential 
\rf{V-eff-cornell} for charged test particles is exclusively due to the presence of
the constant vacuum electric field \rf{const-electr} even though 
Schwarzschild-{\em anti}-de-Sitter is an electrically neutral background.

%%%%%%%%%%%%%%%%%%%%%%%%%%%%%%%%%%%%%%%%%%%%%%%%%%%%%%%%%%%%%%%%%%%%%%%%%%%%%%%%%%
\section{Discussion}
\label{discuss}

It is possible to rewrite the action \rf{gravity+GG} in an explicitly Weyl-conformally
invariant form using the method of two volume forms (two integration measures) 
\ct{TMT} introduced earlier in the context of gravity-matter models with
primary applications in cosmology.
Namely, apart from the standard reparametrization covariant integration
density $\sqrt{-g}$ in terms of the intrinsic Riemannian metric $g_{\m\n}$
as in \rf{gravity+GG}, one introduces an alternative repara\-met\-ri\-zation covariant 
integration density $\P (\vp)$ in terms of auxiliary scalar fields $\vp^I$ 
($I=1,\ldots ,4$):
\be
% \P (\vp) = 1/4! \vareps^{\m_1,\m_2,\m_3,\m_4} \vareps_{I_1 I_2 I_3 I_4}
% \pa_{\m_1}\vp^{I_1} \pa_{\m_2}\vp^{I_2} \pa_{\m_3}\vp^{I_3} \pa_{\m_4}\vp^{I_4} \; .
\P (\vp) = \frac{1}{4!} \vareps^{\k\l\m\n} \vareps_{IJKL}
\pa_{\k}\vp^{I} \pa_{\l}\vp^{J} \pa_{\m}\vp^{K} \pa_{\n}\vp^{L} \; .
\lab{phi-measure}
\ee
Then the following gravity-gauge-field action:
\be
S = \int d^4 x \P (\vp) \Bigl\lb \frac{g^{\m\n} R_{\m\n}(\G)}{16\pi} 
- \frac{f}{2} \sqrt{-F^2}\Bigr\rb - \frac{1}{4} \int d^4 x \sqrt{-g} F^2 \; ,
% \quad, \quad
% F^2 \equiv F_{\k\l} F_{\m\n} g^{\k\m} g^{\l\n} 
% \quad ,\quad F_{\m\n} = \pa_\m A_\n - \pa_\n A_\m
\lab{TMT-gravity+GG}
\ee
where $R_{\m\n}(\G)$ is the Ricci tensor in the first order formalism
(\textsl{i.e.}, function of the affine connection $\G^\m_{\n\l}$),
is explicitly invariant under Weyl-conformal gauge transformations:
\be
g_{\m\n} \to \rho (x) g_{\m\n} \quad ,\quad
\vp^I \to {\bar \vp}^I = {\bar \vp}^I (\vp) \;\;\; \mathrm{such ~that}\;\;\;
\det\Vert \frac{\pa {\bar \vp}^I}{\pa \vp^J}\Vert = \rho (x) \; .
\lab{weyl-conf}
\ee
The original action \rf{gravity+GG} arises as a special gauge-fixed version
of the Weyl-conformally invariant action \rf{TMT-gravity+GG} upon using the
gauge $\P (\vp) = \sqrt{-g}$.

%%%%%%%%%%%%%%%%%%%%%%%%%%%%%
To conclude let us recapitulate the main results in the present note:

(a) The non-Maxwell gauge field dynamics of the nonlinear action $L(F^2)$ in
curved space-time \rf{gravity+GG} produces dynamically a non-zero positive 
cosmological constant;

(b) The coupled gravity-non-Maxwell-gauge-field system (\rf{gravity+GG} or
\rf{gravity+GG+CC}) exhibits asymptotically de Sitter and asymptotically 
{\em anti}-de Sitter static spherically symmetric (charged) black hole solutions 
with an additional constant radial electric field (apart from the Coulomb one); 

(c) Under certain choice of parameters we find a QCD-like confining-type effective 
potential in charged test particle dynamics in the above black hole backgrounds.

%%%%%%%%%%%%%%%%%%%%%%%%%%%%
Furthermore, one can prove the following inverse statement. If we start with
an action $S = \int d^4 x \sqrt{-g} \Bigl( \frac{R(g)}{16\pi} + L(F^2)\Bigr)$ with an
{\em a priori unknown} gauge field lagrangian $L(F^2)$ and demand that
this theory will possess static spherically symmetric solutions of
Reissner-Nordstr{\"o}m-de-Sitter type, with {\em dynamically generated} (via $L(F^2)$)
cosmological constant, then we derive a unique solution 
$L(F^2) = -\frac{1}{4} F^2 - \frac{f}{2}\sqrt{-F^2}$, which was our starting point 
in \rf{gravity+GG}.

%%%%%%%%%%%%%%%%%%%%%%%%%%%%   %%%%%%%%%%%%%%%%%%%%%%%%%%%%
Going back to the non-linear gauge field Eqs.\rf{GG-eqs} we observe that
there exists a more general {\em vacuum} solution of the latter {\em without} the
assumption of staticity and spherical symmetry:
\be
-F^2 = f^2 = \mathrm{const} \quad \Bigl(\mathrm{recall} \;\;
F^2 \equiv F_{\k\l} F_{\m\n} g^{\k\m} g^{\l\n} \Bigr) \; ,
\lab{const-F2}
\ee
which via Eq.\rf{stress-tensor-F} automatically produces an effective positive
cosmological constant:
\be
T^{(F)}_{\m\n} = - \frac{f^2}{4} g_{\m\n} \quad ,\;\; \mathrm{i.e.} \;\;
\L_{\mathrm{eff}} = 2\pi f^2 \; .
\lab{stress-tensor-F-vac}
\ee
Thus, because of the absence of Coulomb field due to \rf{const-F2} and
assuming absence of magnetic field, we obtain the above described 
Schwarzschild-(anti)-de-Sitter \rf{S-AdS+const-electr} or purely Schwarzschild 
solutions with a vacuum electric field, which according to \rf{const-F2} has constant
magnitude $|\vec{E}| = \sqrt{-\h F^2} = \frac{f}{\sqrt{2}}$ but its
orientation is completely arbitrary.
% The implication of the latter property will be
% absence of Schwinger's pair creation in these backgrounds.
In this disordered vacuum, where the electric field with constant magnitute does not 
point in one fixed direction, a test charged particle will not be able
to get energy from the electric field, instead, it will undergo a kind of 
brownian motion, therefore {\em no} Schwinger pair-creation mechanism will take place.

%%%%%%%%%%%%%%%%%%%%%%%%%%%%   %%%%%%%%%%%%%%%%%%%%%%%%%%%%
The present considerations may be extended to higher space-time dimensions
and thus provide a framework to study novel effects
that could appear relevant in the context of {\em TeV} gravity \ct{TeV-grav}
scenarios where nontrivial gauge field effects % (like confinement)
and gravity effects may be of same order.

%%%%%%%%%%%%%%%%%%%%%%%%%%%%%%%%%%%%%%%%%%%%%%%%%%%%%%%%%%%%%%%%%%%%%%%%%%%%%%%%%%
As a final comment we mention two other interesting phenomena triggered by
the gravity/non-linear-gauge-field system \rf{gravity+GG} in the context of
wormhole physics. First, Misner-Wheeler ``charge without charge'' effect \ct{misner-wheeler} 
is known to be one of the most interesting physical phenomena produced by wormholes.
Misner and Wheeler realized that wormholes connecting two asymptotically flat 
space-times provide the possibility of existence of electromagnetically
non-trivial solutions, where {\em without being produced by any charge source} the
flux of the electric field flows from one universe to the other, thus giving the impression 
of being positively charged in one universe and negatively charged in the other universe.

In an accompanying  note \ct{hiding} we found the opposite effect in wormhole physics,
namely, that a genuinely charged matter source of gravity and electromagnetism
may appear {\em electrically neutral} to an external observer. We show in \ct{hiding}
that this phenomenon takes place when coupling the gravity/gauge-field system
\rf{gravity+GG} self-consistently to a codimension-one charged {\em lightlike} brane as a matter
source. The ``charge-hiding'' effect occurs in a self-consistent wormhole
solution of the above coupled gravity/gauge-field/lightlike-brane system
which connects a non-compact ``universe'', comprising the exterior region of
Schwarzschild-de-Sitter black hole beyond the internal (Schwarzschild-type
horizon), to a Levi-Civita-Bertotti-Robinson-type ``universe'' with two
compactified dimensions (\textsl{cf.} \ct{LC-BR})
via a wormhole ``throat'' occupied by the charged
lightlike brane. In this solution the whole electric flux produced by the charged
lightlike brane is expelled into the compactified Levi-Civita-Bertotti-Robinson-type
``universe'' and, consequently, the brane is detected as neutral by an
observer in the Schwarzschild-de-Sitter ``universe''.

The above ``charge-hiding'' solution can be further generalized to a truly
charge-confining wormhole solution \ct{hide-confine} when we couple the 
gravity/gauge-field system \rf{gravity+GG} self-consistently to {\em two} separate 
codimension-one charged {\em lightlike} branes with equal but opposite charges. 
Namely, the latter system possesses a ``two-throat'' wormhole solution where the 
``left-most'' and the ``right-most'' ``universes''
are two identical copies of the exterior region of the neutral Schwarzschild-de-Sitter 
black hole beyond the Schwarzschild horizon,
whereas the ``middle'' ``universe'' is of generalized Levi-Civita-Bertotti-Robinson
``tube-like'' form with geometry $dS_2 \times S^2$ ($dS_2$ being the
two-dimensional de Sitter space). It comprises the
finite-extent intermediate region of $dS_2$ between its two horizons. Both
``throats'' are occupied by the two oppositely charged lightlike branes and the whole
electric flux produced by the latter is confined entirely within the middle
finite-extent ``tube-like'' ``universe''.

%%%%%%%%%%%%%%%%%%%%%%%%%%%%%%%%%%%%%%%%%%%%%%%%%%%%%%%%%%%%%%%%%%%%%%%%%%%%%%%%%%
%% The Appendices part is started with the command \appendix;
%% appendix sections are then done as normal sections
%% \appendix

%% \section{}
%% \label{}

%%%%%%%%%%%%%%%%%%%%%%%%%%%%%%%%%%%%%%%%%%%%%%%%%%%%%%%%%%%%%%%%%%%%%%%%%%%%%%%%%%
%%%%%%%%%%%%%%%%%%%%%%%%%%%%%%%%%%%%%%%%%%%%%%%%%%%%%%%%%%%%%%%%%%%%%%%%%%%%%%%%%%
\section*{Acknowledgments}
E.N. and S.P. are supported by Bulgarian NSF grant \textsl{DO 02-257}.
Also, all of us acknowledge support of our collaboration through the exchange
agreement between the Ben-Gurion University of the Negev (Beer-Sheva, Israel) and
the Bulgarian Academy of Sciences.
%%%%%%%%%%%%%% ADD-NEW
We are grateful to Stoycho Yazadjiev for constructive discussions.
%%%%%%%%%%%%%%
%%%%%%%%%%%%%%%%%%%%%%%%%%%%%%%%%%%%%%%%%%%%%%%%%%%%%%%%%%%%%%%%%%%%%%%%%%%%%%%%%%
%%%%%%%%%%%%%%%%%%%%%%%%%%%%%%%%%%%%%%%%%%%%%%%%%%%%%%%%%%%%%%%%%%%%%%%%%%%%%%%%%%

\end{document}